\documentclass[aip,reprint]{revtex4-1}

\usepackage{amsmath, bm, amsfonts, amssymb}     
\usepackage{graphicx}                       
\usepackage{xfrac}                          
\usepackage{grffile}                        
\usepackage{color,colortbl}
\usepackage{xcolor}

\usepackage[colorlinks, citecolor=blue, urlcolor=blue, linkcolor=blue]{hyperref}

\newcommand{\xhat}{\hat{\bm{x}}}
\newcommand{\yhat}{\hat{\bm{y}}}
\newcommand{\gdot}{\dot{\gamma}}
\newcommand{\sigmay}{\sigma_{\rm y}}

\newcommand{\tauy}{\tau_{\rm y}}
\newcommand{\taus}{\tau_{\rm s}}
\newcommand{\gdotwall}{\dot{\gamma}_\text{wall}}
\newcommand{\gdotbulk}{\dot{\gamma}_\text{bulk}}
\newcommand{\vupper}{V_\text{upper}}
\newcommand{\vlower}{V_\text{lower}}
\newcommand{\vEuler}{v}
\newcommand{\vbar}{\bar{v}}
\newcommand{\yupper}{y_\text{upper}}
\newcommand{\ylower}{y_\text{lower}}
\newcommand{\ynorm}{y_\text{norm}}
\newcommand{\vnorm}{v_\text{norm}}
\newcommand{\vfit}{v_\text{fit}}
\newcommand{\vs}{V_\text{s}}
\newcommand{\vslower}{V_\text{s,lower}}
\newcommand{\vsupper}{V_\text{s,upper}}
\newcommand{\deltahet}{\delta_\text{het}}

\newcommand{\change}[1] {\textcolor{black}{#1}}

\begin{document}
	
	\title{Wall slip and bulk yielding in soft particle suspensions}
	
	\author{Gerhard Jung}
	\affiliation{Institut f\"ur Theoretische Physik, Leopold-Franzens-Universit\"at Innsbruck, Technikerstra{{\ss}}e 21A, A-6020 Innsbruck, Austria}
	\affiliation{Institut f\"ur Physik, Johannes Gutenberg-Universit\"at Mainz, 
		Staudingerweg 9, 55128 Mainz, Germany}
	\affiliation{Department of Physics, Durham University, Science Laboratories,
		South Road, Durham DH1 3LE, UK}
	\author{Suzanne M. Fielding}
	\affiliation{Department of Physics, Durham University, Science Laboratories,
		South Road, Durham DH1 3LE, UK}
	
	\begin{abstract}
		
		We simulate a dense athermal suspension of soft particles sheared between 
		hard walls of a prescribed roughness profile, fully accounting for the 
		fluid mechanics of the solvent between the particles, and for the solid 
		mechanics of changes in the particle shapes. We thus capture the widely 
		observed rheological phenomenon of  wall slip.  For imposed stresses 
		below the material's bulk yield stress, we show the slip to be dominated 
		by a thin solvent layer of high shear at the wall. At higher stresses, it 
		is augmented by an additional contribution from a fluidisation of the 
		first few layers of particles near the wall.  By systematically varying 
		the wall roughness, we quantify a suppression of slip with increasing 
		roughness. We also elucidate the effects of slip on the dynamics of 
		yielding following the imposition of a constant shear stress, 
		characterising the timescales at which bulk yielding arises, and at which 
		slip first sets in.
		
	\end{abstract}
	
	\maketitle
	
	\section{Introduction}
	
	Concentrated suspensions of soft particles, such as
	microgels~\cite{seth2011micromechanical},
	emulsions~\cite{cohen2014rheology}, surfactant
	vesicles~\cite{fujii2006size}, block copolymer
	micelles~\cite{cloitre2011block}, and multi-arm star
	polymers~\cite{vlassopoulos2009polymers}, display both solid and
	liquid rheological (deformation and flow) properties.  Given an
	imposed shear stress lower than some yield stress, $\sigma<\sigmay$,
	they typically show a solid-like creep response in which the shear
	strain $\gamma$ slowly increases over time $t$, but with an ever
	decreasing shear rate, $\gdot\sim t^{-\alpha}$. For a larger imposed
	stress, $\sigma>\sigmay$, an early-time creep regime is followed,
	after a time that is often fit to the form
	$\tauy\propto(\sigma-\sigmay)^{-\delta}$~\cite{divoux2011stress}, by a
	yielding transition to a fluidised state of steady flow with a
	time-independent shear rate $\gdot$, in which $\sigma(\gdot)$ is often
	fit to the `Herschel-Bulkley' form $\sigma=\sigmay+k\gdot^n$, with
	$n\leq 1$~\cite{bonn2017yield}. During yielding, the shear field
	within the fluid bulk often becomes highly
	heterogeneous~\cite{divoux2012yielding,divoux2011stress}.
	
	The motion of such materials is determined not only by their bulk
	properties, however, but also by their interaction with the confining
	walls. For smooth enough walls, a material will often appear to slip
	relative to
	them~\cite{cloitre2017review,barnes1995review,vinogradov1975flow,vinogradov1978generalized,dimitriou2014crudeoil}:
	the velocity profile $v(y)$ across a sheared sample does not meet up
	with the velocity of the walls, but has an apparent mismatch known as
	the slip velocity, $\vs$. This has been suggested to arise via a
	mechanism in which soft particles become deformed by shear and lift
	away from the wall, leaving a thin lubricating solvent layer across
	which a strong shear occurs, giving apparent
	slip~\cite{meeker2004slipJoR,meeker2004slipJoR}. (The hydrodynamic
	no-slip condition is however finally obeyed where the solvent meets
	the wall.) This is thought to be key to numerous processes in nature
	and technology, e.g., water-lubricated
	transport~\cite{joseph1997core}, food transport in the
	gut~\cite{stokes2013oral}, and the squeezing of red blood cells
	through capillaries~\cite{roman2016going}.
	
	A series of remarkable experiments have shown wall slip to have a
	major impact on rheological data, which must be carefully interpreted
	to disentangle the contributions of bulk flow and
	slip~\cite{yoshimura1988wall}. Indeed, slip radically changes the
	steady state flow curve, $\sigma(\gdot)$, by causing a non-zero
	apparent flow branch even below the bulk yield stress,
	$\sigma<\sigmay$~\cite{meeker2004slipJoR,meeker2004slipPRL}. The
	steady state slip velocity $\vs(\sigma)$ typically depends as a power
	law on $\sigma$ or $\sigma-\sigmay$ (below or above $\sigmay$). The
	value and universality of the exponent remain controversial:
	depending on the particle packing fraction and wall properties
	(wetting {\it vs.} non-wetting), experiments report a quadratic
	scaling at small
	stresses~\cite{poumaere2014unsteady,meeker2004slipJoR,meeker2004slipPRL,salmon2003towards,geraud2013confined}
	then linear at larger stresses~\cite{Pemeja2019}, or vice
	versa~\cite{seth2012soft,Gonzalez2012,mansard2014boundary,zhang2017wall,zhang2018wall},
	or a progression from linear to quadratic across an array of
	suspensions from dilute to jammed~\cite{divoux2015wall}.  Very recently, a linear scaling was demonstrated at low stresses, universally across many suspensions and wall types, provided contact line effects are removed~\cite{zhang2017wall,zhang2018wall}, although it is worth noting that the nonlinear scalings of~\cite{geraud2013confined,Pemeja2019} were obtained in microchannels, without edge effects. 
	
	Slip
	also profoundly influences the {\em dynamics} of yielding, during
	which a state of initially solid-like response gives way to a finally
	fluidised
	flow~\cite{divoux2012yielding,divoux2011stress,divoux2010transient,grenard2014timescales,gibaud2009shear,gibaud2010heterogeneous,gibaud2008influence,kurokawa2015avalanche,perge2014time}. Indeed,
	yielding often appears to initiate via slip at the wall, before a
	fluidised band propagates across the bulk to finally fluidise the
	whole sample. The degree of slip is however strongly influenced by
	confinement~\cite{goyon2008spatial,davies2008thin}, wall
	roughness~\cite{mansard2014boundary,barrat2013model} or chemical
	coating~\cite{seth2012soft,seth2008influence,christel2012stick,chan2013simple},
	bringing the intruiging prospect of controlling bulk flows by
	tailoring the wall conditions.
	
	Compared with this remarkable experimental progress, simulation has
	lagged far behind, despite its potentially central role in addressing
	experimentally controversial issues such as the scaling of $\vs$ with
	$\sigma$, and the dependence of $\vs$ on features such as wall
	roughness, which is only rarely varied systematically in
	experiment~\cite{mansard2014boundary}.
	
	The contributions of this paper are fourfold. First, we introduce a
	method of simulating a dense suspension of soft particles sheared
	between hard walls of any prescribed roughness profile. It accounts
	fully for the hydrodynamics of the solvent between the particles, and
	near the walls, and for the elastic solid mechanics via which the soft
	particles change shape. It is thus capable of properly capturing
	rheological wall slip. (Most existing methods instead simply assume a
	spherical interparticle potential and an effective solvent drag, although more advanced methods also
	exist~\cite{RAHIMIAN2010sim,seth2011micromechanical,Derzsi2017,Derzsi2018}.) Second, we quantify the effects
	of slip on steady state flow behaviour, confirming that it radically
	changes a material's flow curve $\sigma(\gdot)$ by conferring a branch
	of slip-induced apparent flow even for $\sigma<\sigmay$. We show that
	the steady state slip velocity $\vs=\nu(\beta)(\sigma-\sigmay)$ for
	$\sigma>\sigmay$, with a transition in which the prefactor $\nu$ drops
	sharply above a critical wall roughness $\beta^*$, suppressing
	slip. For $\sigma<\sigmay$, we separately find $\vs\propto\sigma$ with
	smooth walls. Below yield, slip is indeed dominated by a thin
	Newtonian layer at the wall.  In important contrast, however, above
	yield it additionally includes a fluidisation of the first few layers
	of particles. Third, we elucidate the effects of slip on the {\em
		dynamics} of yielding following the imposition of a constant stress,
	characterising the timescales $\tauy(\sigma)$ at which bulk yielding
	arises, and $\taus(\sigma,\beta)$ at which slip first sets in as a
	material starts to flow.  Finally, we show that slip and bulk effects
	can be disentangled, with master creep and flow curves for the fluid
	bulk, regardless of wall roughness.
	
	The paper is structured as follows. In Sec.~\ref{sec:simulation} we describe our simulation method. Sec.~\ref{sec:parameters_observables} details the physical parameters involved, and the physical observables measured. In Sec.~\ref{sec:results} we present our results and in~\ref{sec:conclusions} give conclusions and suggestions for future work.
	
	\section{Simulation method}
	\label{sec:simulation}
	
	We now introduce our method for simulating a two-dimensional dense
	suspension of soft particles, sheared between hard walls of any
	prescribed roughness profile. Any reader who is not interested in these technical details can jump direct to Sec.~\ref{sec:parameters_observables} without loss of thread.
	
	\subsection{Initialization}
	\label{sec:init}
	
	\subsubsection{Molecular dynamics of circular particles}
	
	{ We take a box of length $L_x$ and height $ L_y $ with periodic boundaries in $x$ and $ y $. Inside the box we randomly initialise an ensemble of $p=1\cdots P$ circular particles in a region of length $L_x$ and height $H-b$ with packing fraction $\phi=0.5$.} (In the next stage, the particles will be expanded to attain a higher $\phi$.) To avoid crystallisation we
	take a bidisperse 50:50 mixture with particle radii in ratio
	$1:1.4$. Particles closer than a distance $ r_{c,pp'}$ interact
	\emph{via} a Lennard-Jones (LJ) potential:
	\begin{align}
	\bm{F}_p^\text{LJ}&= - \bm{\nabla}_p E^\text{LJ}(\{\bm{X}_{p'}\}),\\
	E^\text{LJ}(\{\bm{X}_{p'}\}) &= 4 K_\text{LJ}\sum_{p,p'<p}  \left [ \left (  \frac{\sigma_{pp'}}{X_{pp'}}  \right )^{12} - \left (   \frac{\sigma_{pp'}}{X_{pp'}} \right )^6   \right ].
	\label{eq:LJ_rep}
	\end{align}
	Here $\bm{X}_p$ is the position of the $p$th particle, $X_{pp'} =
	|\bm{X}_{p} - \bm{X}_{p'}| $ the distance between the $p$ and $p'$th
	particles, $K_\text{LJ}$ a force constant and $\sigma_{pp'}$ a
	length.  Each particle also experiences dissipative drag and thermal
	noise, and accordingly obeys (subject to additive corrections from wall interactions to be described in the next paragraph) the equation of motion:
	\begin{equation}
	M\ddot{\bm{X}}_p = \bm{F}_p^\text{LJ} -\frac{M}{\tau} \dot{\bm{X}}_p + \bm{F}_p^{R}.
	\label{eqn:eom}
	\end{equation}
	Here $M$ is the particle mass, $\tau$ a time-constant and $\bm{F}^{R}$
	a delta-correlated random variable with zero mean and variance
	${\frac{k_\text{B} T M}{\Delta t
			\tau}} $.  
	
	Parallel walls are placed above and below the particle packing a
	distance $\Delta y = H$ apart. Each comprises a flat line of length
	$L_x$, periodically interrupted by semicircular bumps of radius $b$
	and separation $B$, which protrude into the packing.  Each wall is
	discretised into many ($N_w$) nodes (we shall return below to discuss
	the value of $N_w$), and each wall node is held in a fixed position. A
	short-ranged LJ force then additionally acts between the particles and
	wall nodes. This is of the form of Eqn.~\ref{eq:LJ_rep}, with the
	particle labels $p'$ augmented by wall node labels $s'$. An overview of the parameter values in this molecular dynamics stage is shown in Table~\ref{tab:one}.
	
	{To remove  particle-particle  and particle-wall node overlaps,
		we first minimize the interaction energy using the
		Polak-Ribiere version of the conjugate gradient algorithm
		(provided by the LAMMPS package \cite{Plimpton1995}). } The
	equations of motion, Eqn.~\ref{eqn:eom}, are then temporally
	discretised using the Velocity-Verlet
	algorithm~\cite{vel_verlet} and evolved using
	LAMMPS~\cite{Plimpton1995}
	\footnote{\url{https://lammps.sandia.gov/}} with a time step $\Delta t$
	until the ensemble reaches a statistical steady state after a time
	$\tau_\text{eq} = 5000 \Delta t$.
	
	\begin{table}[!b]
		\begin{tabular}{|c|c|c|}
			\hline 
			Symbol & Parameter & Value \\ 
			\hline 
			$P$ & number of particles & $400$  \\ 
			\hline 
			$L_y$ & box height & $1.0$ [length unit] \\ 
			\hline 
			$L_x$ & box width & $0.5$  \\ 
			\hline 
			$\phi$ & area packing fraction & $0.5$  \\ 
			\hline 
			$M$ & particle mass & $1.0$ [mass unit]  \\ 
			\hline 
			$K_\text{LJ}$ & LJ energy constant & 1.0 [energy unit] \\ 
			\hline 
			$\sigma_{pp'}$ & LJ length constant (particle-particle) & $1.2 (R_p+R_{p'})$  \\ 
			\hline 
			$r_{c,pp'}$ & LJ cutoff length (particle-particle) & $\sqrt[6]{2}\sigma_{pp'}$ \\ 
			\hline 
			$\tau$ & Langevin time constant & 0.01 \\ 
			\hline 
			$T$ & temperature & 0.1 \\ 
			\hline 
			$\Delta t$ & numerical time step & $5.42\times 10^{-6}$ \\ 
			\hline 
			$H$ & wall separation & $0.44$   \\
			\hline 
			$b$ & wall bump radius & [varied]  \\
			\hline 
			$B$ & wall bump separation & $5.0b$  \\ 
			\hline 
			$\sigma_{ps'}$ & LJ length constant (particle-wall) & $1.2 R_p $ \\ 
			\hline 
			$r_{c,ps'}$ & LJ cutoff (particle-wall) & $\sqrt[6]{2}\sigma_{ps'}$ \\ 
			\hline 
		\end{tabular} 
		\caption{Parameter values in the molecular dynamics stage.}
		\label{tab:one}
	\end{table}
	
	\subsubsection{Particle expansion and shape change}

	\begin{table}[b]
		\hspace*{-0.8cm}	\begin{tabular}{|c|c|c|}
			\hline 
			Symbol & Parameter & Value \\ 
			\hline 
			$N_{s1}$ & boundary nodes per smaller particle &  250  \\ 
			\hline 
			$K_\text{e}$ & particle boundary elastic constant & 2.0 [2 $\times$ energy unit]\\ 
			\hline 
			$K_\text{p}$ & expansion force constant & 0.5 \\ 
			\hline 
			$\sigma_{s s^\prime}$ & LJ length constant & $0.00125 $  \\ 
			\hline 
			$r_{c,s s^\prime}$ & LJ cutoff & $\sqrt[6]{2}\sigma_{s s^\prime}$ \\ 
			\hline 
			$K_\text{LJ}$ & LJ energy constant & 0.01 \\ 
			\hline 
			$\gamma$ & drag & 1.0 [sets time unit] \\ 
			\hline 
			$\Delta t$ & numerical time step & $1.125\times 10^{-6}$ \\ 
			\hline 
		\end{tabular} 
		\caption{Parameters values in the particle expansion stage. Values for $P$, $ L_x $, $L_y$, $H$, $b$, $B$, [length unit] as in Table.~\ref{tab:one}.}
		\label{tab:two}
	\end{table}
	
	After the molecular dynamics equilibration just described, the
	boundary of each (initially) circular particle is discretised into
	evenly distributed surface nodes, separated a distance (initially) of
	$\Delta s=2\pi R/N_s$. (We therefore use two different values of $N_s$,
	in ratio $1:1.4$, to ensure the same $\Delta s$ for the two particle
	species.)
	The particle boundaries are then rendered elastic via a
	force acting between adjacent nodes round each boundary according to an elastic membrane model \cite{peskin_2002}. The continuous version of this model is given by,
	\begin{align}
	\bm{F}^\text{elastic}(s) &= \frac{\partial}{\partial s} (T \bm{\tau}), \nonumber\\
	\bm{\tau} &= \frac{\partial \bm{X}/\partial s}{|\partial \bm{X}/\partial s|}, \nonumber\\
	T&= K_e \left( \frac{\partial \bm{X}}{\partial s} - 1 \right),
	\end{align}
		with boundary tension $ T $, unit tangent $ \bm{\tau} $ and $K_e$ a surface elastic force constant. This force is discretised to calculate the force on each boundary node,
	\begin{align}
	\bm{F}^\text{elastic}_{s} &= \frac{T_{s+1/2}\bm{\tau}_{s+1/2}-T_{s-1/2}\bm{\tau}_{s-1/2}}{\Delta s},\nonumber\\
	\bm{\tau}_{s+1/2} &= \frac{\bm{X}_{s+1} - \bm{X}_{s}}{|\bm{X}_{s+1} - \bm{X}_{s}|},\nonumber\\
	T_{s+1/2}&= K_e \left ( \frac{|\bm{X}_{s+1} - \bm{X}_{s}|}{\Delta s} -1 \right ),
	\label{eq:elastic_force}
	\end{align}
	The index $s=0\cdots
	N_s-1$ runs over the nodes of any particle boundary, with periodic
	boundary conditions. For clarity we omit here the particle number
	label $p$. Note that the actual distance $|\bm{X}_{s+1} - \bm{X}_{s}|$
	between any two nodes will change during the simulation, whereas the
	equilibrium distance remains constant and equal to $\Delta s$.
	
	As noted above, parallel walls are located above and below the
	particle packing a distance $\Delta y = H$ apart.  Each wall is
	discretised into $N_w$ nodes, with neighbouring nodes
	separated by the same curvilinear distance $\Delta s$ that (initially)
	separates neighbouring nodes in the particle boundaries. (Accordingly,
	the actual number $N_w$ used in any simulation depends on the values
	of $b$ and $B$.)  The wall nodes remain fixed in position during this part of the simulation.
	As above, a short-ranged LJ force acts between the
	nodes of different particles, and between particle and wall
	nodes. This is of the form of Eqn.~\ref{eq:LJ_rep}, with the particle
	labels $p'p$ augmented by node labels $s's$.
	
	The particles are expanded by a pressure that acts inside each particle,
	modelled via a force of amplitude $K_p$ acting on each boundary node
	along the outward normal:
	\begin{equation}
	\bm{F}^\text{pressure}_{s} = K_p(\bm{\tau}_s' \times \bm{\hat{z}}),
	\end{equation}
	with centred tangent $\bm{\tau}_s' = (\bm{X}_{s-1} -
	\bm{X}_{s+1})/|\bm{X}_{s-1} - \bm{X}_{s+1}| $. The boundary and wall nodes  move as $\dot{\bm{X}}_s=\bm{F}_s/\gamma$, where $\bm{F}_s$ is the total force on any node, against a drag $\gamma$, without explicit hydrodynamics in this initialisation phase. This equation is evolved using the explicit Euler algorithm with time step $\Delta t$. As they expand, the particles change shape due to crowding, but avoid overlap via the short-ranged LJ potential.  The wall shapes remain constant, with particle-wall overlaps also avoided by the LJ potential. The simulation is stopped when the desired area fraction is achieved. An overview of the parameter values in this particle expansion stage is shown in Table~\ref{tab:two}.
	
	\subsection{Shearing with hydrodynamics}
	\label{shearing}

	The configuration of particle boundary and wall nodes attained at the
	end of the initialisation procedure just described is then transferred
	to form the initial configuration in a code that now also incorporates
	shearing and hydrodynamics. In this hydrodynamic code, the Lagrangian
	particle boundary and wall nodes move relative to a fixed rectangular
	Eulerian mesh, on which we implement incompressible Stokes flow. The
	combined Eulerian and Lagrangian dynamics are handled using an
	``immersed boundary method''~\cite{peskin_2002}. At any time step, this comprises
	the following substeps:
	\begin{enumerate}
		\item Given the source forces arising from the current configuration of particle boundary and wall nodes, as mapped onto the Eulerian mesh in step 5 of the previous time step, the Stokes equation is solved to find the fluid velocity on the Eulerian mesh.
		\item This Eulerian  velocity field is mapped onto the Lagrangian particle boundary and wall nodes.
		\item Using these Lagrangian velocities, the Lagrangian positions of the particle boundary nodes are updated. From  these new positions,  the new Lagrangian forces of the particle boundary nodes are calculated.
		\item Again using the Lagrangian velocities from step 2, the Lagrangian positions and forces of the wall boundary nodes are updated.
		\item These forces on the Lagrangian particle boundary and wall nodes are mapped to the Eulerian mesh.
	\end{enumerate}
	Each substep is detailed in the correspondingly numbered subsection
	below.
	
	At the start of the shearing simulation, we set the desired
	equilibrium distance $\Delta s$ between adjacent boundary nodes of
	each particle to be equal to $L/N_s$, where $L$ is the perimeter of a
	circle with the same area as that particle. This desired equilibrium
	distance then remains constant during the shearing simulation that
	follows.  Additionally, the wall nodes are initialised with forces
	$\bm{F}_s= (+\sigma L_x/(N_w\Delta s),0)$ and $ (-\sigma
	L_x/(N_w\Delta s),0)$ on the upper and lower walls respectively, to
	impose a shear stress on the soft particle suspension. The algorithm
	that follows then keeps this shear stress constant over the course of
	the simulation.
	
	\begin{table}[b]
		\hspace*{-0.4cm}
		\begin{tabular}{|c|c|c|}
			\hline 
			Symbol & Parameter & Value \\
			\hline 
			$\eta$ & solvent viscosity & $1.0$ [viscosity unit]  \\ 
			\hline 
			$K_e$ & particle surface elastic constant & $1.0$ [stress unit]  \\ 
			\hline 
			$\phi$ & particle area fraction & $0.84$  \\
			\hline 
			$R$ & average particle radius & $0.0085$  \\ 
			\hline 
			$\sigma_\text{LJ}$ & LJ length constant & $9 \text{d}x$  \\ 
			\hline 
			$r_{c}$ & LJ cutoff & $\sigma_\text{LJ}$ \\ 
			\hline 
			$K_\text{LJ}$ & LJ energy constant & $ 10^{-9} $ \\ 
			\hline 
			$N_x, N_y$ & number of Eulerian grid points & $ 4096 L_x, 8192  L_y $ \\ 
			\hline 
			$\alpha = \Delta s / \text{d}x $ & Lagrangian/Eulerian grid ratio & $  1.42 $ \\ 
			\hline 
			$K_w$ & wall elastic constant & $ 20000 $ \\ 
			\hline 
			$\Delta t$ & numerical time step & $0.002$ \\ 
			\hline 
		\end{tabular} 
		\caption{Parameters used in shearing stage. Values for $P$, $ L_x $, $L_y$ [length unit], $H$, $B$, $b$, $N_s$, as in Table.~\ref{tab:two} and/or~\ref{tab:one}.}
		\label{tab:parameters}
	\end{table}
	
	\subsubsection{Stokes flow on the Eulerian mesh}
	
	We consider a biperiodic plane of size $L_x\times
	L_y$ in which are located Lagrangian walls a distance $\Delta y=H$
	apart. These will move relative to each other in order to perform
	shear. In the gap of size $H$ are soft particles and a Newtonian
	solvent of viscosity $\eta$. In the complementary gap of size $L_y-H$
	there is Newtonian solvent only.  Over the full $L_x\times L_y$ plane,
	the fluid velocity field $\bm{v}(\bm{x},t)$ and pressure field
	$p(\bm{x},t)$ obey the incompressible Stokes equations:
	\begin{align}
	0 &=  \eta \nabla^2 \bm{v} + \bm{f}- \bm{\nabla} p,\\
	0 &= \bm{\nabla} \cdot \bm{v}.
	\label{Stokes_equation}
	\end{align}
	Here $\bm{f}(\bm{x},t)$ is a source force density field, which acts
	only at the walls of the shearing cell, and round the boundaries of
	the soft particles.  These Stokes equations are discretised on a fixed
	rectangular Eulerian mesh of $i=0\cdots N_x-1,j=0\cdots N_y-1$ points,
	with the same mesh size $\text{d}x=L_y/N_y=L_x/N_x$ in both $x$ and
	$y$. (We describe below how to map the wall and particle boundary
	Lagrangian forces onto this Eulerian mesh.) The discretised
	differential operator is defined as:
	\begin{equation}
	D_x \phi_{i,j} = \frac{\phi_{i+1,j}-\phi_{i-1,j}}{2 \,\text{d}x},
	\end{equation}
	for any discretised field $\phi_{i,j}$, with $D_y \phi_{i,j}$ defined
	similarly. The discretised Stokes equations are then:
	\begin{align}
	0 &=  \eta \mathcal{D}^{2} \bm{u}_{i,j} + \bm{f}_{i,j} - \bm{D}\, p_{i,j},\label{eqn:Dstokes} \\ 
	0 &= \bm{D} \cdot \bm{u}_{i,j},
	\end{align}
	with $\bm{D} = (D_x,D_y,0)^T$ and $\mathcal{D}^2 = \bm{D}
	\cdot \bm{D} $.  
	
	We enforce the incompressibility condition by introducing a
	streamfunction $\Psi(\bm{x},t)$ via $\bm{v}_{i,j} = \bm{D} \times
	(\Psi_{i,j} \bm{\hat{z}})$, and eliminate the pressure by taking the
	curl of Eqn.~\ref{eqn:Dstokes}:
	\begin{align}
	0 = - \eta \mathcal{D}^4\Psi_{i,j} + (\bm{D} \times \bm{f}_{i,j})\cdot \bm{\hat{z}}.\label{eqn:singleStokes}
	\end{align}
	This equation can then be solved using fast Fourier transforms (FFT)
	with a computational cost that scales as $N_y N_x \ln(N_x N_y)$. 
	
	The discrete FT is defined as
	\begin{equation}
	\hat{\phi}_{k_x,k_y} = \sum_{i=0}^{N_x-1} \sum_{j=0}^{N_y-1} e^{-i(2\pi/N_x) ik_x} e^{-i(2\pi/N_y) j k_y} \phi_{i,j}.
	\end{equation}
	The FT of Eqn.~\ref{eqn:singleStokes} is:
	\begin{equation}
	0 = - \eta \frac{16}{\text{d}x^4} \left [ \sin^2\left(\frac{\pi k_x}{N_x}\right)+\sin^2\left(\frac{\pi k_y}{N_y}\right)  \right ]^2 \hat{\Psi}_{k_x,k_y} + \hat{\mathcal{F}}_{k_x,k_y},
	\end{equation}
	where 
	$\hat{\Psi}_{k_x,k_y}$ and $\hat{\mathcal{F}}_{k_x,k_y} $ are the FTs of $\Psi_{i,j}$ and $(\bm{D} \times \bm{f}_{i,j})\cdot \bm{\hat{z}}$ respectively.
	For any source force field $\bm{f}_{i,j}$, this equation is solved to
	find the FFT of the stream function, $\hat{\Psi}_{k_x,k_y}$. Via the
	inverse FFT, we find finally the streamfunction $\Psi_{i,j}$ and fluid
	velocity $v_{i,j}$ on the Eulerian mesh.
	
	We define by $\alpha\equiv \Delta s/\text{d}x$ the ratio of the parameter
	$\Delta s$, which we recall sets the separation of Lagrangian
	mesh points, and the mesh size $dx$ of the Eulerian grid. The value of
	this parameter is important to the effectiveness of any immersed
	boundary simulation. Too large a value will lead to fluid leakage across the particle
	boundaries \cite{BAO2017183}. Too small a value leads to an increased computational
	effort. Throughout we use a value $\alpha=1.42$.

	\subsubsection{Eulerian to Lagrangian velocity mapping}
	
	The discretised velocity field $\bm{v}_{i,j}$ as calculated on the
	Eulerian mesh in the previous substep is now interpolated to the
	Lagrangian particle boundary and wall nodes using the formula:
	\begin{equation}
	\bm{V}_s =\sum_{i=0}^{N_x - 1}\sum_{j=0}^{N_y - 1} \bm{v}_{i,j} \delta_h(\bm{x}_{i,j} - \bm{X}_s)\text{d} x^2.
	\end{equation}
	Here we use a smoothed discretised delta function $\delta_h(\bm{x})=\delta_h(x)\delta_h(y)$~\cite{Yang2009} in which:
	\begin{equation}
	\delta_h(x) = \left\{\begin{array}{ll}
	\frac{3}{8}+\frac{\pi}{32}-\frac{x^2}{4 }, & \text{for } 0 < \frac{|x|}{dx}\leq 0.5 \\
	\frac{1}{4}+\frac{1-|x|}{8}\sqrt{-2+8|x|-4x^2}&\\-\frac{1}{8}\arcsin(\sqrt{2}(|x|-1)), & 
	\text{for } 0.5 < \frac{|x|}{dx}\leq 1.5 \\
	\frac{17}{16}-\frac{\pi}{64}-\frac{3 |x|}{4}+\frac{x^2}{8 }&\\+\frac{|x|-2}{16}\sqrt{-14+16|x|-4x^2}&\\+\frac{1}{16}\arcsin(\sqrt{2}(|x|-2)), & \text{for } 1.5 < \frac{|x|}{dx}\leq 2.5 \\
	0, & \text{for } \frac{|x|}{dx} >  2.5.
	\end{array} \right. 
	\label{eq:peskin_delta}
	\end{equation}
	\begin{figure}
		\includegraphics[width=8.00cm]{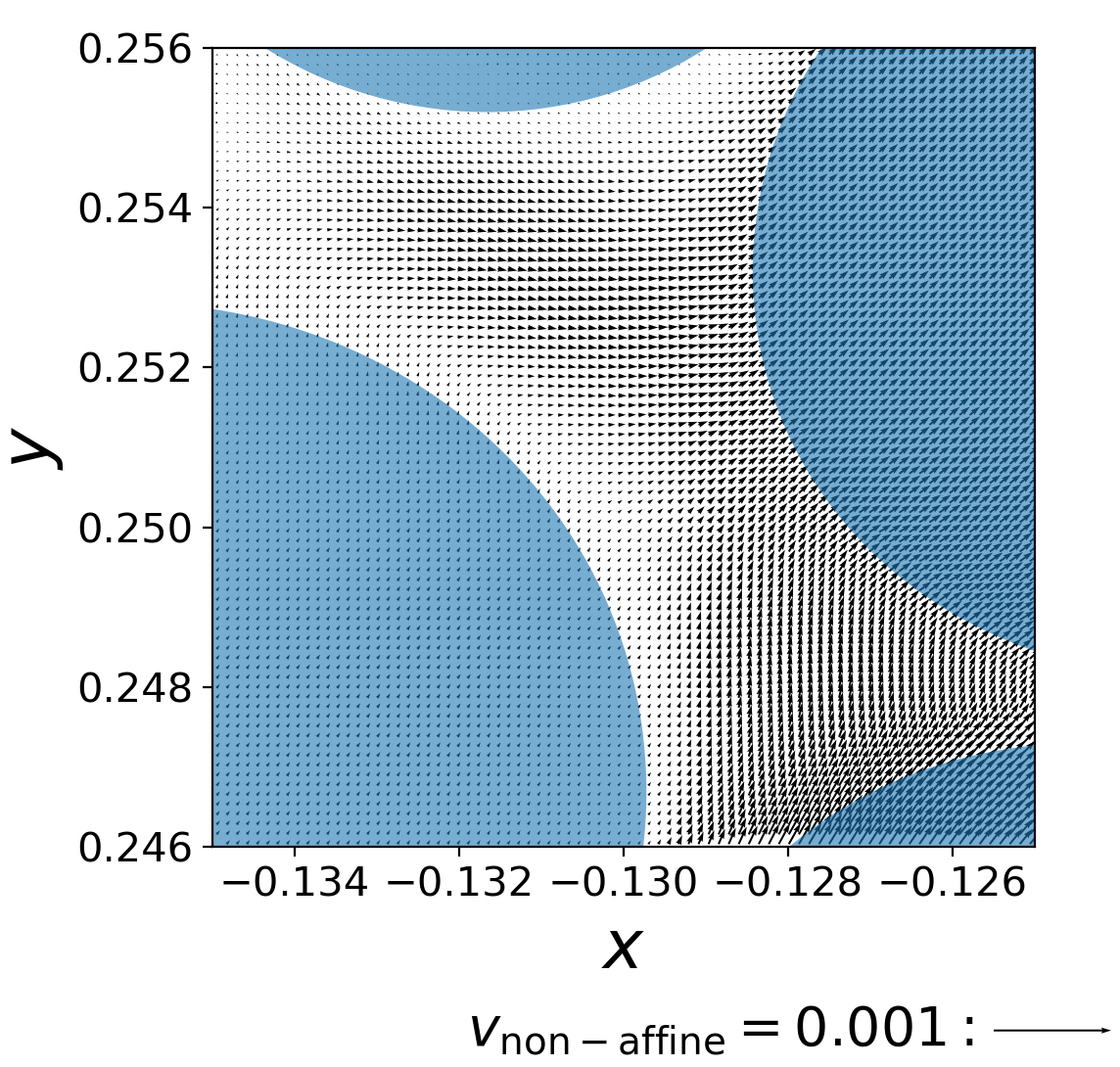}
		\caption{ Simulation snapshot showing the full resolution of the hydrodynamic velocity field. Snapshot taken at time $t=254.0$, imposed stress $\sigma=0.3$. Soft particles (blue), inter-particle fluid (white). Black arrows show non-affine velocity field $v_{\rm na}$, as defined in Eqn.~\ref{eq:na_velocity}. } 
		\label{fig:snapshot2}
	\end{figure}
	
	A snapshot of the soft particles and the velocity field on the Eulerian mesh is shown in Fig.~\ref{fig:snapshot2}. As can be seen, the method is able to fully resolve the hydrodynamic velocity field in the channels, including no slip boundary conditions and long-range hydrodynamic interactions.
	
	\subsubsection{Lagrangian dynamics of the particle boundary nodes}
	
	Given the Lagrangian velocities of the particle boundary nodes as just
	calculated, their positions can in principle be updated from time step
	$n\to n+1$ simply by using an explicit Euler algorithm:
	\begin{equation}
	\bm{X}_s^{n+1} = \bm{X}_s^n + \Delta t \,\bm{V}_s.
	\label{eqn:Euler1}
	\end{equation}
	For clarity, we omit  here any particle number label $p$ from $\bm{X}_s$, and
	include only node label $s$.
	
	With such an update, the area of each particle should in principle
	remain constant due to the incompressibility of Stokes flow. In
	practice, however, using the raw $\bm{V}_s$ in Eqn.~\ref{eqn:Euler1}
	gives a small error in particle area conservation due to fluid leakage
	across the particle boundary. Over an entire simulation this was
	about $1\%$ in the worst case. To correct for this, we used the
	following method~\cite{Newren2007}.
	
	Strict particle area conservation requires that over the area $\Omega$
	and boundary $\partial\Omega$ of each particle:
	\begin{align}
	\int_{\Omega} \bm{\nabla} \cdot \bm{v} \,\text{d} A = \int_{\partial \Omega} \bm{v} \cdot \bm{n}\, \text{d} S =0,
	\end{align}
	where we have used the divergence theorem in writing the first equality.
	In discretized form this reads:
	\begin{equation}
	0 = \sum_{s=0}^{Ns-1} \bm{V}_s \cdot \bm{\hat{n}}_s \,\Delta S_s,
	\label{eq:constraint_dis}
	\end{equation}
	with $\bm{\hat{n}}_s = \bm{{n}}_s/|\bm{{n}}_s|$, $\bm{{n}}_s =
	(Y_{s-1} - Y_{s+1} ,X_{s+1} - X_{s-1},0)^T$, and $\Delta S_s =
	|\bm{{n}}_s|/2.0$. To enforce this constraint we define
	\begin{equation}
	M = \sum_{s=0}^{N_s-1} \bm{V}_s \cdot \bm{\hat{n}}_s \Delta S_s/\sum_{s=0}^{N_s-1} \Delta S_s,
	\end{equation}
	and subtract this mean value from the normal velocity of any particle boundary node:
	\begin{equation}
	\bm{V}_s \rightarrow \bm{V}_s - M \bm{\hat{n}}_s.
	\end{equation}
	We use this corrected velocity in the explicit Euler update.  With
	this, the worst case variation in any particle area over a full
	simulation is smaller than $0.1\%$.
	
	Given the updated $\bm{X}_s$ round the boundary of each particle, the
	elastic boundary forces $\bm{F}^\text{elastic}_{s}$ are then
	recalculated using Eqn.~\ref{eq:elastic_force}. (In this, recall that
	the value of the equilibrium internode length $\Delta s$ is a constant
	and equal to its value as at the start of the shearing simulation.)
	
	The nodes of different particles also interact via a weak, truncated LJ
	force $\bm{F}^\text{LJ}_{s}$ of the same general form as in
	Eqn.~\ref{eq:LJ_rep}. This force  introduces a new length scale, $\sigma_\text{LJ}$, which corresponds, for example, to the physics of a van der Waals interaction. \change{The interaction length scale of the LJ potential was empirically adjusted such that the particles never get so close that the finite discretisation of the Lagrangian nodes becomes a limitation in the hydrodynamic solver. It therefore also ensures that no diverging lubrication forces emerge.}   Particle nodes also
	interact with the wall nodes in the same way. The potential used is
	now of slightly softer form, however, with
	\begin{align}
	E^\text{LJ}(\{\bm{X}_i\}) &= 4 K_\text{LJ} \sum_{s,s'<s  }  \left ( 3 \left (  \frac{\sigma_\text{LJ}}{X_{ss'}}  \right )^{8} - 4\left (   \frac{\sigma_\text{LJ}}{X_{ss'}} \right )^6   \right ).
	\end{align}

	The total force on any particle boundary node is then
	$\bm{F}_s=\bm{F}^\text{elastic}_{s}+\bm{F}^\text{LJ}_{s}$.

	\subsubsection{Lagrangian dynamics of the wall nodes}
	
	\begin{figure}
		\includegraphics[width=9.00cm]{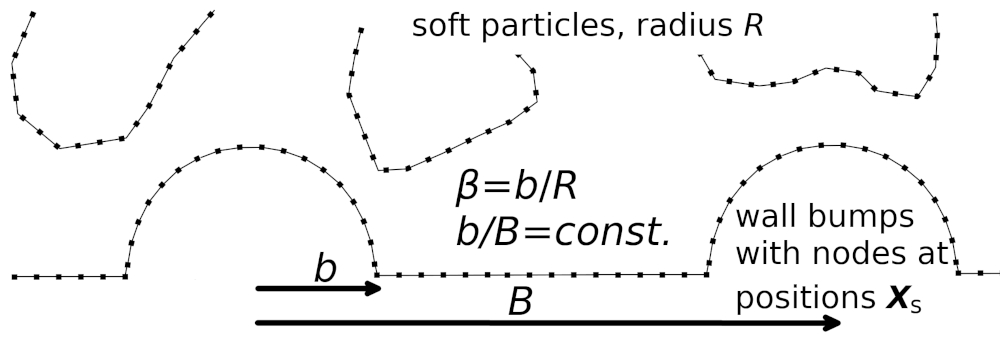}
		\caption{ Schematic of the (lower) wall. The wall is composed of individual nodes at positions $ \bm{X}^\text{lower}_s $. The flat parts of the walls are interrupted by regular bumps (semi-circles) of radius $ b $ separated by a distance $B$. } 
		\label{fig:schematic_wall}
	\end{figure}
	
	A schematic of the wall is shown in Fig.~\ref{fig:schematic_wall}. The wall itself is modelled as stiff, with the relative distance between wall nodes kept constant using a similar approach as in Eqs. (15),(16) of  Refs.~\cite{HEMINGWAY2018viscoelastic}. In the following we describe how to generalize this approach to constant-stress simulations in a channel.
	
	To effect a relative shearing motion of the walls under conditions of
	a constant imposed shear stress, any node $s$ of the upper wall is
	tethered at any time $t$ in the simulation via a strong spring force
	to a desired moving position $\bm{X}^\text{upper}_s(0)+\xhat
	\int_0^t dt
	\hat{V}^\text{upper}(t)$, where $\bm{X}^\text{upper}_s(0)$ was that
	node's initial location. Accordingly we have:

	\begin{align}
	&\bm{F}^\text{tether,upper}_s(t)=	\label{eqn:wall1}\\&-K_w\left[ \bm{X}^\text{upper}_s(t)- \left(\bm{X}^\text{upper}_s(0)+\xhat \int_0^t dt' \hat{V}^\text{upper}(t')\right) \right]. \nonumber
	\end{align}

	Here $\hat{V}^\text{upper}(t)$ is the time-dependent rightward speed
	of the upper wall that must obtain (along with a counterpart leftward
	speed $\hat{V}^\text{lower}(t)$ for the lower wall, described below)
	in order to maintain a constant imposed shear stress in the soft particle
	suspension between the walls. Our aim in what follows is to calculate
	these required wall speeds.  Averaging Eqn.~\ref{eqn:wall1} across all
	nodes in the upper wall, and denoting the average by an overbar,
	gives:

	\begin{align}
	&	L_x\sigma\xhat + \frac{L_x}{L_y-H}\eta(\hat{V}^\text{upper}+\hat{V}^\text{lower})\xhat + G^\text{upper}\yhat\nonumber\\
	=&-K_w\left[ \bar{\bm{X}}^\text{upper}_s(t)- \left(\bar{\bm{X}}^\text{upper}_s(0)+\xhat \int_0^t dt \hat{V}^\text{upper}(t)\right) \right].
	\label{eqn:wall2}
	\end{align}

	The terms on the LHS arise from area-integrating the force balance
	condition over a rectangle of length $L_x$ that entirely encloses the
	upper wall. (Recall that force balance states that the divergence of
	the stress tensor, plus any body forces, must everywhere equal zero.) 
	Converting this area integral to a surface integral via the divergence
	theorem then gives terms arising from the integral of the shear stress
	separately along the upper and lower boundaries of that rectangle. The
	integrals along the side walls of the rectangle, $ G^\text{upper} $, cancel by virtue of
	the periodic boundary conditions. The upper boundary of the rectangle
	lies in the solvent outside the walls, with the term in $\eta$ giving
	the known shear stress in that Newtonian linear shear profile. (Here
	we have assumed that the semi-circular wall bumps, which are small on
	the scale of the channel height $ L_y-H $, have negligible effect on
	the known result for the shear stress for Newtonian flow between flat
	parallel walls.) The lower boundary lies in the soft particle packing between
	the walls. Here we define $\sigma$ to be the $x-$averaged shear stress
	in that packing, which must be independent of $y$ across the packing.
	
	Taking the time-derivative of the previous two  equations gives respectively:
	\begin{equation}
	\dot{\bm{F}}^\text{tether,upper}_s(t)=-K_w\left[ \bm{V}^\text{upper}_s(t)- \xhat \hat{V}^\text{upper}(t) \right].
	\label{eqn:wall3}
	\end{equation}
	and (writing now only the $x$ component):
	\begin{equation}
	\frac{L_x\eta(\dot{\hat{V}}^\text{upper}+\dot{\hat{V}}^\text{lower})}{L_y-H} 
	=-K_w\left[ \bar{V}^\text{upper}_{x}(t)- \hat{V}^\text{upper}(t) \right].
	\label{eqn:wall4}
	\end{equation}

	Note that the time-derivative of the shear stress $\sigma$ across the
	packing, which would appear in Eqn.~\ref{eqn:wall4}, is zero in this
	constant-stress simulation. Exactly corresponding counterparts to Eqns.~\ref{eqn:wall1}
	to~\ref{eqn:wall4} can then be written for the lower wall.

	We calculate the Lagrangian velocities $\bm{V}_s$ of the wall nodes in step 2 above, and thus we can determine their $x-$components averaged
	separately across all nodes forming the upper and lower walls,
	$\bar{V}^\text{upper}_{x}(t)$ and $\bar{V}^\text{lower}_{x}(t)$. Therefore,
	Eqn.~\ref{eqn:wall4} and its counterpart for the lower wall form two
	coupled ordinary differential equations in the desired wall
	speeds, $\hat{V}^\text{upper}(t)$ and $\hat{V}^\text{lower}(t)$, that
	must be imposed to maintain a constant shear stress within the
	suspension. We update these imposed wall speeds by stepping these ODEs
	via the explicit Euler algorithm with time step $\Delta t$.
	
	These updated imposed wall speeds $\hat{V}^\text{upper}(t)$ and
	$\hat{V}^\text{lower}(t)$, together with the wall node velocities
	$\bm{V}_s$ as calculated in step 2, are then substituted into
	Eqn.~\ref{eqn:wall3} and its counterpart for the lower wall, which are
	used to update the tether forces $\bm{F}^\text{tether}_s$ on the wall
	nodes, again using the explicit Euler algorithm with a time step
	$\Delta t$.
	
	The velocities $\bm{V}_s$ of the wall nodes are also used to update
	the positions of the wall nodes. In principle, we should perform the
	update using the velocity of each node separately: $\bm{X}_s^{n+1} =
	\bm{X}_s^n + \Delta t \,\bm{V}_s$. However, over the course of a
	simulation this can lead to a small deformation in the shape of each
	wall. We therefore instead use the average node velocity for each
	wall. Therefore, for all nodes in the upper wall we compute
	\begin{equation}
	\bm{X}_s^{n+1} = \bm{X}_s^n + \Delta t \,\bar{\bm{V}}^\text{upper},
	\label{eqn:Euler2a}
	\end{equation}
	with a corresponding expression for the lower wall.

	\subsubsection{Lagrangian to Eulerian force mapping}
	
	The Lagrangian forces on the particle boundary and wall nodes are
	finally mapped onto the Eulerian mesh. For any particle or either
	wall we perform the sum:
	\begin{equation}
	\bm{f}_{i,j} = \sum_{s=0}^{N_s - 1} \bm{F}_s \delta_h(\bm{x}_{i,j} - \bm{X}_s) \Delta s,
	\label{eq:discrete_mapping}
	\end{equation}
	(with $N_s$ replaced by $N_w$ for the walls), further summing over all
	particles and both walls.  Here we use the same discretised delta
	function as adopted above in Eq.~(\ref{eq:peskin_delta}) in mapping the Eulerian velocities to the
	Lagranian nodes.
	
	\section{Physical parameters and observables}
	\label{sec:parameters_observables}
	
	\begin{figure}
		\includegraphics[width=8.00cm]{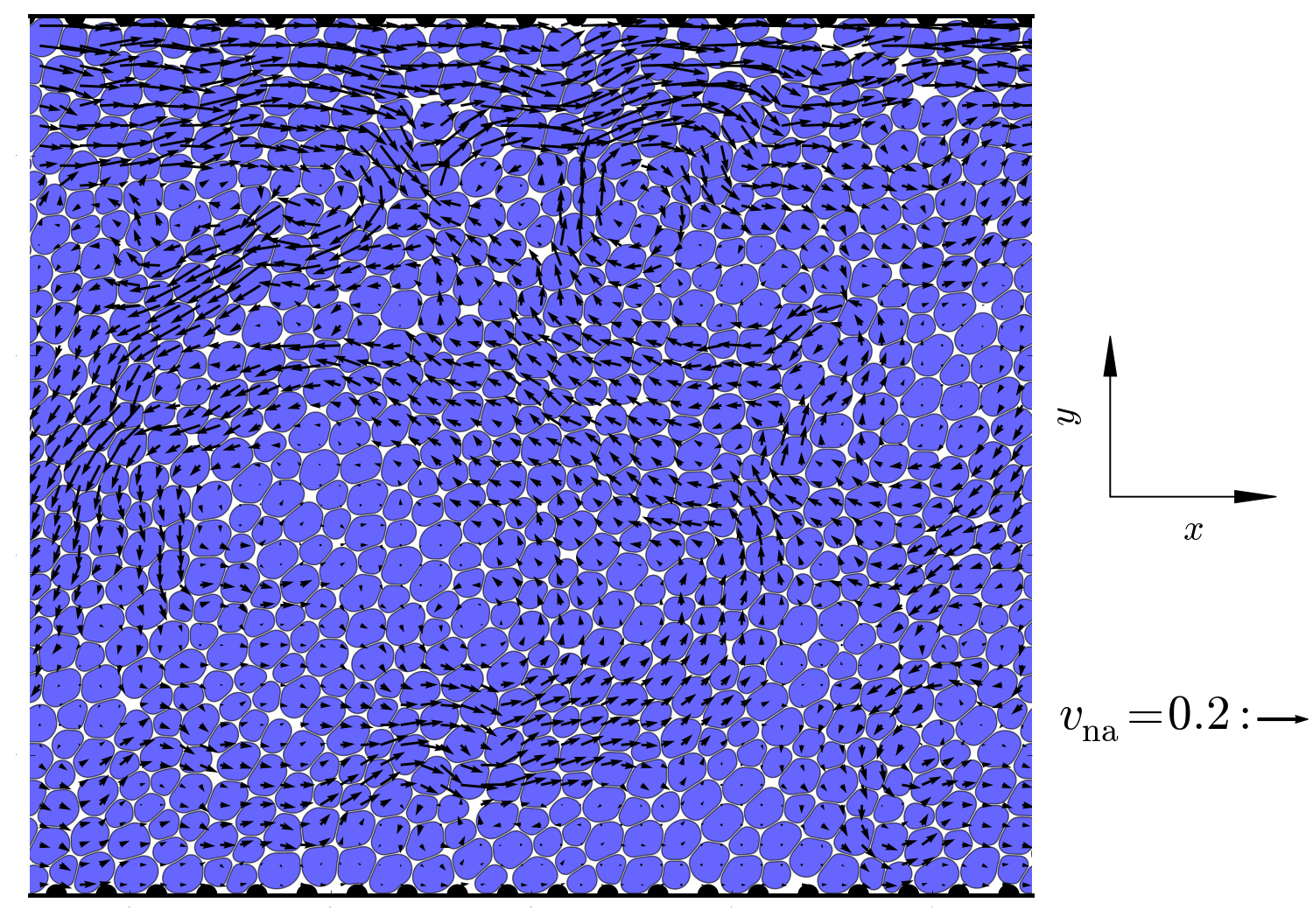}
		\caption{Simulation snapshot at time $t=17.5$, as the sample yields. Soft particles (blue), inter-particle fluid (white), rough hard walls (black). Arrows show non-affine velocity field $v_{\rm na}$. Wall roughness $\beta = 0.59$, imposed stress $\sigma=0.5$. 
		}
		\label{fig:snapshot}
	\end{figure}
	
	The simulation parameters are as follows: the height of the simulation
	box, $L_y=1.0$ (length unit); the height of the channel containing the
	sheared soft particles, $H=0.44$ (the space of height $L_y-H$ outside
	the channel contains only Newtonian solvent); the channel length,
	$L_x=0.5$; the radius, $b$, and separation, $B$, of the wall bumps,
	which we keep in fixed ratio $B/b=5.0$ across all runs; the number of
	soft particles $N=800$; the particle boundary elastic constant
	$K_e=1.0$ (stress unit); the wall elastic constant $K_{\rm
		wall}=20000.0$; the solvent viscosity $\eta=1.0$ (viscosity unit); the
	LJ parameters between nodes of neighbouring particles; and
	the numerical time step and mesh parameters. The particle
	area fraction is fixed at $\phi=0.84$ (giving the average particle
	radius $R=0.0085$). Combined with the repulsive part of the LJ
	potential, this ensures the packing is jammed at rest.  Parameters to
	be explored are then the scaled wall roughness $\beta\equiv b/R$ and
	imposed shear stress $\sigma$.

	\begin{figure*}[!t]
		\includegraphics[scale=1.5]{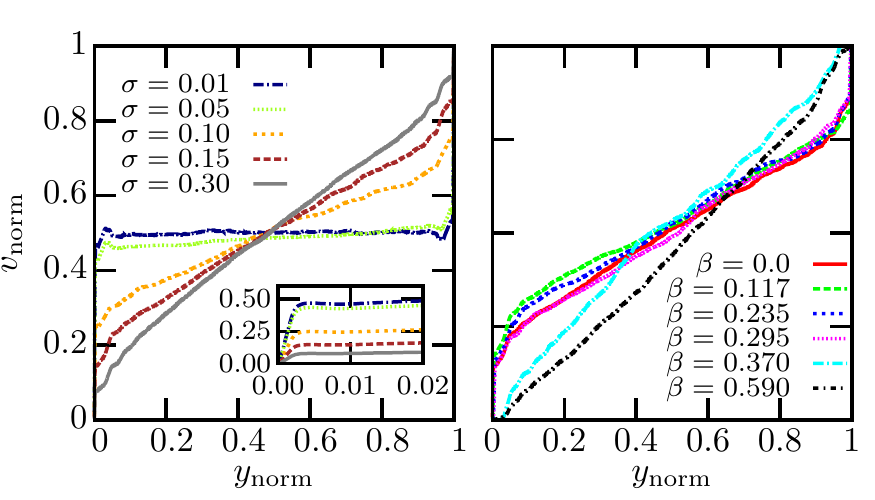}
		\caption{Steady state velocity profiles. {\bf Left)}:  at
			several shear stresses for smooth walls, $\beta=0.0$. Inset: zoom near wall. {\bf Right}):  several wall roughnesses at shear stress $\sigma =0.15$.}
		\label{fig:profiles}
	\end{figure*}
	
	We measure from our simulations the Lagrangian wall velocities
	$\vlower$ and $\vupper$. The apparent shear rate across the channel is
	then $\gdotwall=(\vupper-\vlower)/H$. This includes a contribution
	from true shear across the fluid bulk, and from a thin slip layer
	adjacent to each wall. To disentangle these contributions, we measure
	the flow speed in the main flow direction $x$ at any location on the
	Eulerian grid between the walls as $\vEuler(x,y)$, and average it
	along $x$ to get the velocity profile $\vbar(y)$ across the
	channel. Over a reduced gap that excludes the slip layers, from
	$y=\ylower+5R$ to $y=\yupper-5R$, we fit $\vbar(y)$ to a straight
	line, $\vfit(y)$. The slope of this line then gives the bulk shear
	rate $\gdotbulk$, and its wall intercepts give the slip velocities:
	$\vslower=\vfit(y=\ylower)-\vlower$ and
	$\vsupper=\vupper-\vfit(y=\yupper)$. We report the average slip
	velocity $\vs=(\vslower+\vsupper)/2$. We define the normalised
	velocity profile $\vnorm(y) = (\vbar(y)-\vlower)/(\vupper-\vlower)$
	versus $\ynorm(y)=(y-\ylower)/H$. We have checked
	that our results for $\gdotbulk$ and $\vs$ show no finite size
	dependence on $H$ (see Appendix~\ref{app:fs}). $\gdotwall$ of course does depend on $H$, due to
	the important effect of slip itself. Indeed, this is how slip was
	measured experimentally~\cite{yoshimura1988wall}, before the use of flow velocimetry. We also define the non-affine velocity \begin{equation}\label{eq:na_velocity}
	{\bm{v}}_\text{na}(x,y) = \tfrac{1}{\gdotbulk H}\left[\bm{v}_\text{Euler}(x,y)-v_\text{fit}(y)\,\xhat\right]
	\end{equation}
	and characterise the flow heterogeneity in the fluid bulk (over the reduced gap $ y_\text{lower}+5R < y' < y_\text{upper}-5R $) as \begin{equation}\label{key}
	\delta_\text{het}=\tfrac{\sqrt{\Lambda}}{\dot{\gamma}_\text{bulk}H\sqrt{N_xN_{y'}}} 
	\end{equation}
	with $\Lambda=\sum_{x,y'} [v_\text{Euler,x}(x,y') -v_\text{fit}(y')]^2$. 
	
	We also analyse two distinct slip lengths. The Newtonian slip length $ l^\text{newton}_s $ describes the thickness of a channel with purely Newtonian flow directly at the wall. It is defined as the point of largest curvature in the tangential velocity profile $ \bar{v}(y) $ close to the wall, $ \text{d}^3\bar{v}(y)/\text{d}y^3|_{y=l^\text{newton}_s}=0 $.
	The total slip length $ l^\text{tot}_s $ describes the distance of the wall to an extrapolated point in space for which the tangential velocity component vanishes (corresponding to the typical definition of slip length). This characterises the full slip layer, which includes the Newtonian layer just described as well as the first few layers of particles near the wall, which experience an increase in fluidisation.
	
	Any steady state quantity reported in this
	work is averaged in each run between the time $t_{\rm ss}$ at which it
	visibly attains a steady state, and $t_{\rm ss} +
	\Delta t$ where $\Delta t >= 500.0$. Each is further averaged over at
	least two independent simulations. The error bars then correspond to
	the standard error in the distribution of the time-series averages
	across these independent simulations.
	
	A sample particle packing, including the rough boundary, is shown in Fig.~\ref{fig:snapshot} to give the reader a visualization of the system. 
	
	\section{Results}
	\label{sec:results}

	\begin{figure*}
		\includegraphics[scale=1.6]{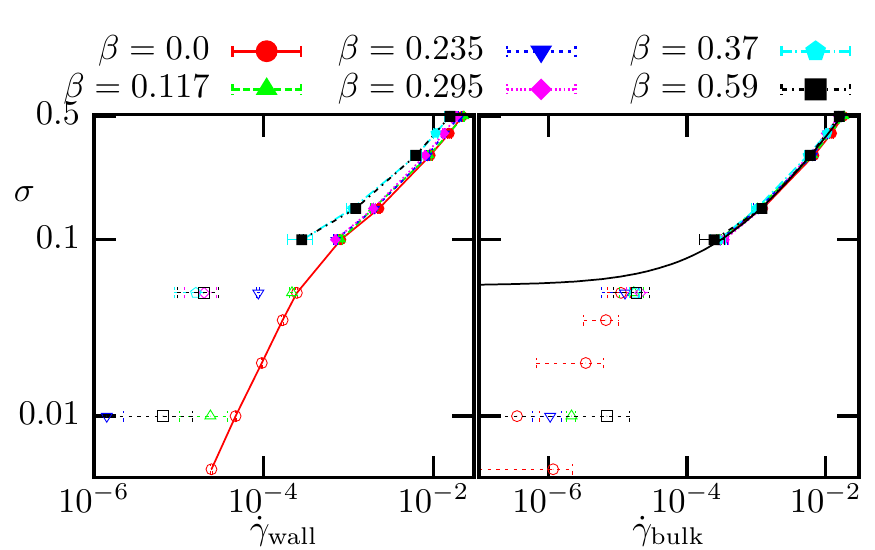}	
		\caption{{\bf Left)} Apparent flow curves with shear rates calculated from relative wall speeds, including wall slip, for different wall roughnesses $ \beta $. Solid line connects data points for smooth wall case. {\bf Right)} Corresponding bulk flow curves using shear rate obtained from internal velocity profile, with slip removed.  Filled symbols: steady state (errors bars too small to be seen at high stresses). Unfilled symbols: do not attain steady state, with dashed error bars showing drift during the time {$t>2000$} over which data is taken. Solid line: fit to $ \sigma=\sigmay+k\gdotbulk^n$ with $ \sigmay = 0.055 \pm 0.004 $ and $ n = 0.57 \pm 0.03 $.}
		\label{fig:flow_curve}
	\end{figure*}

	\subsection{Steady-state velocity profiles and flow curves}

	The steady state normalised profiles
	(Fig.~\ref{fig:profiles}, left) reveal two separate contributions to
	the slip: one from a very thin solvent layer within about $\Delta y =
	0.0025$ of the wall (inset), and another over about $\Delta y = 0.1$,
	corresponding to an increase in fluidity over the first few particle
	layers near the
	wall~\cite{mansard2014boundary,Derzsi2017,derzsi2018wall}.
	Importantly, we find the first contribution to dominate the total slip
	at stresses below yield, whereas above yield both are important.  We
	report the total slip, because it is more likely to be the one seen in
	experimental velocimetry of realistic pixel resolution.  We note that
	$\vbar(\ylower)=\vlower$ (as seen in the inset) and
	$\vbar(\yupper)=\vupper$, consistent with hydrodynamic no-slip for the
	solvent.
	
	Fig.~\ref{fig:flow_curve} shows the steady state flow curve
	relationship between the imposed shear stress $\sigma$ and the shear
	rate $\gdot$, for several different values of the wall roughness
	parameter $\beta$. (Although in our simulations $\sigma$ is imposed
	and $\gdot$ measured, we show $\sigma(\gdot)$ because this is the
	usual flow-curve representation.)  The left panel has as its abscissa
	the apparent shear rate, $\gdotwall$, defined via the relative wall
	speed. As noted above, this includes not only any true shear across
	the fluid bulk, but also the effects of wall slip. The right panel
	uses the true bulk shear rate, $\sigma(\gdotbulk)$, with slip
	removed. Above a yield stress, $\sigma>\sigmay$, the steady state data
	superpose for all levels of wall roughness, once slip is removed. The
	resulting flow curve is then fit to the Herschel-Bulkley form,
	$\sigma=\sigmay+k\gdot^n$, with $\sigmay=0.055\pm 0.004$ and
	$n=0.57\pm 0.03$. For lower stresses, $\sigma<\sigmay$, $\gdotbulk$
	does not attain a steady state, as indicated by the open symbols in
	Fig.~\ref{fig:flow_curve}.

	\begin{figure*}[!t]
		\includegraphics[scale=1.6]{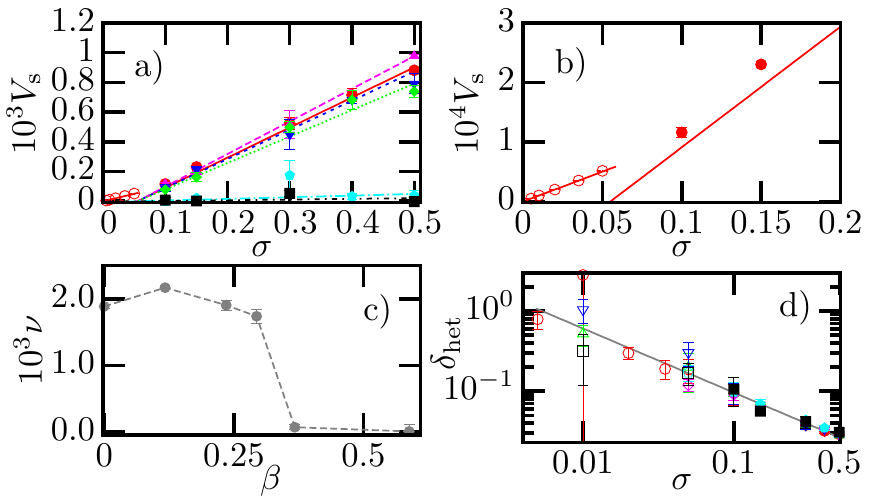}
		\caption{{\bf a)} Symbols: steady state slip velocity vs. imposed stress for different wall roughnesses, with  roughness symbol legend as in  Fig.~\ref{fig:flow_curve}. Lines:  least-square  fits to  $ \vs(\sigma> \sigma_\text{y}) = \nu(\beta) (\sigma - \sigma_\text{y}) $ and $ \vs(\sigma < \sigma_\text{y}) = \nu_\text{N}(\beta) \sigma$.  {\bf b)}: Zoom of $\beta=0.0$ data for  $\sigma<0.2$. {\bf c)}: Prefactor $\nu$ vs. wall roughness $ \beta $.	{\bf d)}: Steady state degree of heterogeneity in the bulk flow field vs. imposed stress for different surface roughnesses. Solid line: $\deltahet\propto \sigma^{-0.8}$ as a guide to the eye. 
		}
		\label{fig:steady_state_slip}
	\end{figure*}

	\subsection{Heterogeneous flow profiles: Wall slip and bulk yielding}
	
	We now further explore the extent to which the flow profiles across
	the gap are heterogeneous due to wall slip and non-affine flows in
	the bulk. Fig.~\ref{fig:steady_state_slip}a) shows the steady state
	wall slip velocity as a function of imposed shear stress, for several
	levels of wall roughness, $\beta$. The data for $\sigma>\sigmay$ are
	fit for each roughness to the form $\vs=\nu(\beta)(\sigma-\sigmay)^p$,
	with $p=1$. We also find $p=1$ with an essentially unchanged
		$\nu(\beta)$ if we instead allow a free intercept, $ \sigma_Y^\prime $. This linear
	dependence for $\sigma>\sigmay$ is consistent with the experiments of Ref.~\cite{seth2012soft,mansard2014boundary} whereas those of Ref.~\cite{salmon2003towards,geraud2013confined}
	found a quadratic dependence, $p=2$. In Ref.~\cite{divoux2015wall}, it was suggested that $p$ varies
	between $1$ and $2$ as a function of packing fraction $\phi$ above
	jamming. It would be interesting in future to simulate a range of
	$\phi$. The prefactor $\nu$, plotted as a function of $\beta$ in panel
	c), reveals a transition from strong slip for smooth walls, with
	$\beta<\beta^*\approx 0.3$, to suppressed slip for rougher walls,
	$\beta>\beta^*$. A decreasing slip with increasing wall roughness was
	seen for wall roughnesses less than the average particle size
	($\beta<1$) in Ref.~\cite{mansard2014boundary}.
	
	For $\sigma<\sigmay$, we find a different scaling of the slip
	velocity, $\vs\propto\sigma$, for smooth walls. (For rough walls,
	$\vs$ takes prohibitively long to attain a steady state.) That we
	observe different scalings for $\vs(\sigma)$ above and below yield is
	consistent with the discussion above regarding
	Fig.~\ref{fig:profiles}, left: that slip below yield is dominated by a
	thin solvent layer at the wall, with an additional contribution above
	yield from  fluidisation of the first few particle layers. 
	
	The transition between the two scalings, below and above yielding, appears to be rather sharp, but a smoother transition is possible within the error bars, which would allow for a small window in which the exponent $ p > 1 $.
	
	\begin{figure}
		\includegraphics[scale=1.00]{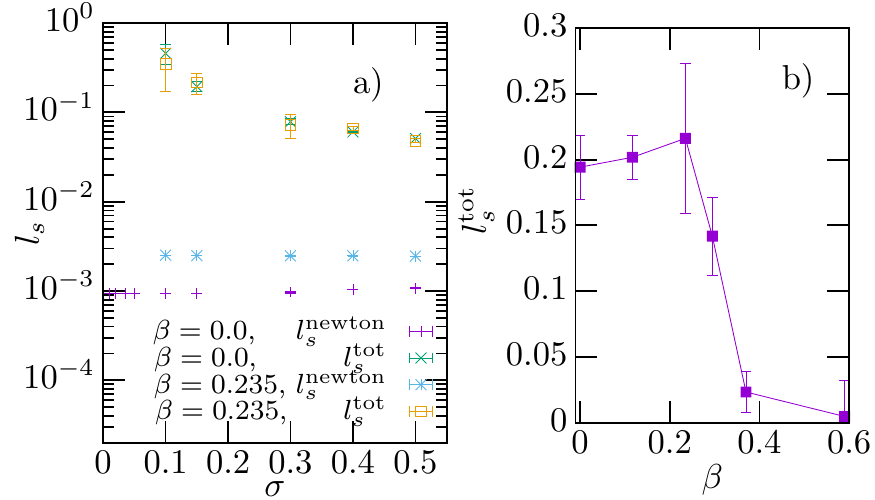}
		\caption{Steady-state slip length $ l_s $ for different external stresses $ \sigma $ (a) and wall roughness $ \beta $ (b). The Newtonian slip length $ l^\text{newton}_s $ describes the thickness of the thin Newtonian layer at the wall and the total slip length $ l^\text{tot}_s $ describes the distance of the wall to an extrapolated point in space for which the tangential velocity component vanishes. } 
		\label{fig:slip_length}
	\end{figure}
	
	The wall slip can be further characterised using the Newtonian slip length, $ l^\text{newton}_s $, and the total slip length $ l^\text{tot}_s $ as defined in Sec.~\ref{sec:parameters_observables}. We observe that the Newtonian slip length is approximately independent of the applied stress $ \sigma $ (see Fig.~\ref{fig:slip_length}a).  The length scale corresponds to the range of the Lennard-Jones interaction between the particles, $\sigma_\text{LJ}$, plus the bump size $ b = \beta R $. This indicates that, first, the precise nature of the direct particle-particle and particle-wall significantly influences slip and, second, that below the ``critical'' roughness $ \beta < 0.35 $, despite a bump size significant larger than the particle-wall interactions, a Newtonian slip layer emerges. For $ \sigma > 0.2 $ and $ \beta=0 $ (corresponding to a flat wall) we additionally observe that the particles lift further from the wall than for small stresses (roughly $ 20\% $ for $ \sigma = 0.5 $) which could be connected to the process of hydrodynamic lift described in Ref.~\cite{meeker2004slipPRL}. Contrary to the Newtonian slip length, the total slip length $ l^\text{tot}_s $ does not depend on surface roughness (below $ \beta < 0.35 $) instead it strongly depends on the applied stress $ \sigma $. This can be rationalized by the difference in scaling of the slip velocity $ V_s $ (linear) and the true shear rate $ \dot{\gamma}_\text{bulk} $ (super-linear). Additionally, the total slip length $ l^\text{tot}_s $ displays the same discontinuous transition from slip to no-slip that was discussed for $ V_s $ (see see Fig.~\ref{fig:slip_length}b).

	In addition to this apparent slip at the walls, the flow profile
	within the fluid bulk also shows strong departures from affine
	shear. This is already apparent in the snapshot of
	Fig.~\ref{fig:snapshot}, in which the {arrows} show the degree to
	which the flow velocity at any location differs from a simple linear
	shear profile. In Fig.~\ref{fig:steady_state_slip}d) we quantify the
	bulk flow heterogeneity (on average, in steady state) via the
	parameter $\deltahet$, plotted as a function of the imposed stress for
	several different levels of wall roughness. For imposed stresses $ \sigma < \sigma_Y $, both $ \sqrt{\Lambda} $ and $ \dot{\gamma}_\text{bulk} $ which appear in the definition of $ \deltahet $, are very small and do not attain a steady state, as indicated by the open symbols (therefore the large error bar). We have, however, observed that the value of the flow heterogeneity itself is stationary during creep, which enables the calculation of a meaningful average. The results clearly indicate an increase of the heterogeneity with
	decreasing imposed stress as $\deltahet\sim\sigma^{-0.8}$. It is
	relatively independent of wall roughness, showing that the effects of
	the wall persist only a few particle diameters into the bulk.  This
	result suggests that the dynamical heterogeneity diverges at $\sigma
	\rightarrow 0 $ under  conditions of imposed stress, distinct
	from the divergent avalanche size seen at low imposed strain rate
	$\gdot\to 0$~\cite{lemaitre2009rate}.
	
	\begin{figure*}
		\includegraphics[scale=1.5]{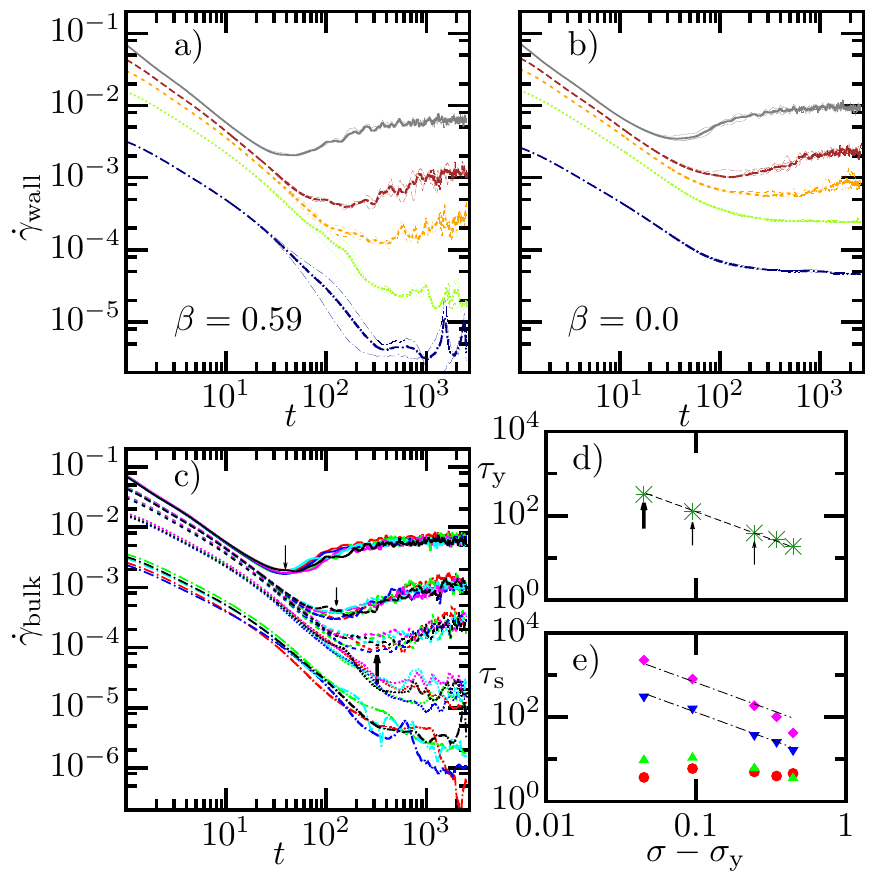}
		\caption{Apparent shear rate vs. time for shear stresses $ \sigma = 0.3,0.15,0.1,0.05,0.01 $ in curve bundles downwards for {\bf (a)} a rough wall and {\bf (b)} a smooth wall. (In each bundle, thick line shows average over 2 or 4 runs; thin lines show individual runs.)
			{\bf (c)} Corresponding true shear rate vs. time for the same imposed
			stresses. (In each bundle, curves are for several roughnesses, with
			colour code as in Fig.~\ref{fig:flow_curve}. For each roughness, curve
			is averaged over 2 or 4 runs.)  {\bf (d)} Yielding time $\tauy$ at the
			minimum in $\gdotbulk(t)$ (averaged over roughnesses), as a function
			of stress above yield. Dotted line: power $-1.3$. (Arrows denoting times in c) and d) coincide.) {\bf (e)} Time
			$\taus$ at which wall slip velocity $\vs$ attains half its
			steady state value for the 4 smoothest walls, with roughness symbols
			as in Fig.~\ref{fig:flow_curve}. 
			Dot-dashed line:
			power $-1.3$.  }
		\label{fig:time_dependence}
	\end{figure*}

	\subsection{Transient dynamics and creep curves}

	We now investigate the transient evolution as a function of the time
	$t$ since the imposition of a constant stress $\sigma$ on a sample
	that is freshly prepared then aged for a waiting time $t_{\rm
		w}=50.0$, before shearing starts at $t=0$. In particular, we explore
	the dynamical yielding process via which a regime of initial creep,
	with a strain rate that decreases over time, gives way to a final
	steady state flow.

	Figs.~\ref{fig:time_dependence}a,b) show the apparent shear rate (as
	measured via the relative wall speeds, and so including the effects of
	slip) as a function of time $t$ for a rough wall (a) and smooth wall
	(b). In each case, data are shown for five imposed stress values in separate curve bundles. The highest three stress values are
	all above the yield stress, $\sigma>\sigmay=0.055$. Here, the apparent
	shear rate $\gdotwall$ initially decreases as function of time, before
	attaining a minimum. The sample then yields and the shear rate
	increases to attain a steady flowing state. For the two lowest stress
	values, for which $\sigma<\sigmay$, the apparent shear rate attains a
	steady state only for smooth walls. This is due to the confounding
	effects of slip: with rough walls, where slip is suppressed and
	$\gdotwall$ coincides with the true bulk shear rate $\gdotbulk$, the
	shear rate never attains a steady state but continues to slowly
	decrease.

	For the same five values of stress, the true bulk shear rate is shown
	as a function of time in Fig.~\ref{fig:time_dependence}c). The curve
	bundle for each stress value now shows results for the six values of
	wall roughness explored in the flow curves of
	Fig.~\ref{fig:flow_curve}. Now that the effects of wall slip have been
	removed by plotting $\gdotbulk(t)$, the data for all wall roughnesses {essentially}
	coincide. In this way, we find the yielding dynamics in the fluid bulk
	to be largely independent of wall roughness. We extract by eye the
	time at the minimum in $\gdotbulk$ and define this to be the time
	$\tauy$ at which yielding occurs. This shows a good fit to $\tau\sim
	(\sigma-\sigmay)^{-1.3}$ (Fig.~\ref{fig:time_dependence}d). Similarly,
	we determine the time $\taus$ at which slip first arises at the wall
	(defined as the time at which $\vs(t)$ attains half its steady state
	value). We plot this as a function of $\sigma-\sigmay$ in
	Fig.~\ref{fig:time_dependence}e) for the four lowest values of wall
	roughness explored in the flow curves of
	Fig.~\ref{fig:flow_curve}. (For the two roughest walls in
	Fig.~\ref{fig:flow_curve}, no appreciable slip arises.) This slip
	timescale increases with increasing wall roughness. For the largest
	two roughness values at which slip occurs, $\taus$ further appears to
	depend on stress in the same way as the timescale for bulk yielding,
	with $\taus\sim(\sigma-\sigmay)^{-1.3}$. Whether slip pre-empts bulk
	yielding (or vice versa), as determined by the prefactor, however
	depends on the roughness.
	
	\section{Conclusions}
	\label{sec:conclusions}
	
	To summarise, we have introduced a method for simulating the dynamics
	of a dense athermal suspension of soft particles sheared between hard
	walls of any roughness profile, in order to study the key
	rheological phenomenon of wall slip. For imposed stresses below the
	bulk yield stress, we have shown wall slip to be dominated by a thin
	solvent layer adjacent to the wall. In contrast, for imposed stresses
	above yield we find an additional slip contribution arising from a
	fluidisation of the first few particles layers. We have further characterised the scaling of slip velocity with imposed stress, and demonstrated a transition from strong to suppressed slip with
	increasing wall roughness. We have also characterised the
	dependence of the timescale for yielding within the bulk fluid on the
	imposed stress, and of the timescale for slipping at the wall as a
	function of both imposed stress and wall roughness. In future, it
	would be interesting to study slip in less concentrated soft
	suspensions, below jamming; rougher wall profiles to address a return
	of slip for roughnesses exceeding the particle
	diameter~\cite{mansard2014boundary}; and different interaction
	potentials with the wall.
	
	Since this manuscript was written, we have become aware of a manuscript
	studying the effects of wall slip on a dense suspension of droplets in
	steady state pressure driven flow along a channel~\cite{SucciPreprint}. It
	focuses entirely on steady state behaviour, presenting results for the
	mass throughput along the channel as a function of wall shear stress
	and wall roughness.
	
	\section*{Acknowledgements} We thank Thibaut Divoux for discussions.  This work was funded by the Deutsche Forschungsgemeinschaft (DFG, German Research Foundation) - Project number 233630050 - TRR 146.

	\appendix
	
	\section{Finite Size Effects}
	\label{app:fs}
	
\begin{figure}
		\includegraphics[scale=1.0]{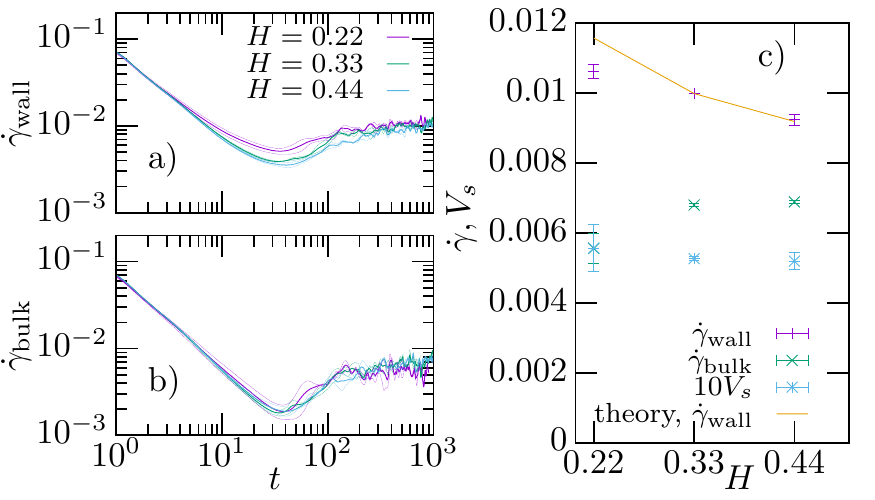}
		\caption{Dependence on wall separation $ H $ of a) the wall shear rate $ \dot{\gamma}_\text{wall} $, b) the true shear rate $ \dot{\gamma}_\text{bulk} $ and c) steady-state values of the former two quantities and the slip velocity $ V_s. $ The orange 'theory' curve is defined as $ \dot{\gamma}_\text{wall}(H) = 2 V_s(H=0.44)/H + \dot{\gamma}_\text{bulk}(H=0.44). $ Definitions of $ \dot{\gamma}_\text{wall}$, $ \dot{\gamma}_\text{bulk}$ and $ V_s$ are given in the main text. } 
		\label{fig:fs}
	\end{figure}

	Since collective plastic rearrangements as well as long-range hydrodynamics can lead to many-body interactions that span several particle diameters, one can expect substantial finite size effects if the wall separation of the channel is too small. In Fig.~\ref{fig:fs} we show results for the transient evolution of strain rate as a function of the time $t$ since the imposition of a constant stress $\sigma$ for different wall separations $ H $. The apparent shear rate $ \dot{\gamma}_\text{wall} $ indeed shows significant finite size effects (see panel a). This result is anticipated because it consists of two contributions,
	\begin{equation}\label{eq:gamma_wall}
	\dot{\gamma}_\text{wall}(H) = 2 V_s/H + \dot{\gamma}_\text{bulk}, 
	\end{equation}
	including the true shear rate in the bulk $\dot{\gamma}_\text{bulk}$ (see panel b), and the slip velocity, $V_s$, both of which are assumed to not depend on the wall separation. The latter assumption is investigated by comparing the steady-state values of the three quantities (see panel c), and indeed no significant deviation can be found for $ H > 0.3 $ (it seems that the bulk flow for $ H=0.22 $ is slightly slower than calculated for large channels). Using Eqn.~(\ref{eq:gamma_wall}) we can show that the dependence of $ \dot{\gamma}_\text{wall} $ on wall separation can actually be accounted to the trivial dependence on the (inverse) wall separation. Inverting this argument highlights a straightforward method to determine the slip velocity. Indeed, this is how slip was measured before the development of advanced experimental techniques like flow velocimetry \cite{yoshimura1988wall}.
	
	\bibliography{slip}
	
\end{document}